\documentclass{article}
\usepackage{PRIMEarxiv}
\usepackage{times}

\usepackage[utf8]{inputenc} 
\usepackage[T1]{fontenc}    
\usepackage{hyperref}       
\usepackage{url}            
\usepackage{booktabs}       
\usepackage{amsfonts}       
\usepackage{nicefrac}       
\usepackage{microtype}      
\usepackage{lipsum}
\usepackage{fancyhdr}       
\usepackage{graphicx}       
\graphicspath{{media/}}     

\usepackage{amsmath}
\usepackage{colortbl}
\usepackage{natbib}

\makeatletter
\renewcommand{\title}[1]{\gdef\@title{#1}} 
\makeatother

\makeatletter
\title{{rquest: An R package for hypothesis tests and confidence intervals for quantiles and summary measures based on quantiles}}
\makeatother

\author{
  Luke A. Prendergast \\
  Department of Mathematical and Physical Sciences \\
  La Trobe University \\
  Melbourne, Australia 3086\\
  \texttt{luke.prendergast@latrobe.edu.au} \\
  \And
     Shenal Dedduwakumara \\
  Faculty of Health and Medical Sciences \\
  University of Adelaide \\
  Adelaide, Australia 5055\\
   \And
  Robert G. Staudte \\
  Department of Mathematical and Physical Sciences \\
  La Trobe University \\
  Melbourne, Australia 3086\\
}

\begin{document}
\maketitle

\begin{abstract}
Sample quantiles, such as the median, are often better suited than the sample mean for summarising location characteristics of a data set.  Similarly, linear combinations of sample quantiles and ratios of such linear combinations, e.g. the interquartile range and quantile-based skewness measures, are often used to quantify characteristics such as spread and skew.  While often reported, it is uncommon to accompany quantile estimates with confidence intervals or standard errors.  The rquest package provides a simple way to conduct hypothesis tests and derive confidence intervals for quantiles, linear combinations of quantiles, ratios of dependent linear combinations (e.g., Bowley's measure of skewness) and differences and ratios of all of the above for comparisons between independent samples.  Many commonly used measures based on quantiles are included, although it is also very simple for users to define their own. 
 Additionally, quantile-based measures of inequality are also considered.  The methods are based on recent research showing that reliable distribution-free confidence intervals can be obtained, even for moderate sample sizes.  Several examples are provided herein.
\end{abstract}


\section{Introduction} 

Our objective is to provide researchers with a comprehensive package that can be used for both one-sample and two-sample analyses (e.g. hypothesis testing, confidence intervals etc.) of statistical measures based on quantiles.  This includes for single quantiles (e.g., the median), linear combinations of quantiles (such as the interquartile range), ratios of linear combinations commonly found in skewness and kurtosis measures, and newly developed inequality measures.  Additionally, while we have sought to include many common measures used in practice, another key objective is to make it easy for users to define their own measures for hypothesis testing and  confidence intervals.

Good quantile estimates have been available on R \citep{R} for decades, with theoretical support provided by \cite{hyndman1996sample}.
Also, there is a substantial literature on methods for estimating the so-called \textit{quantile density function} \citep[see, e.g.,][]{falk1986estimation, welsh1988asymptotically} which is a component of the asymptotic variance of quantile estimators.  \cite{prst-2016a}
find distribution-free standard errors for quantile estimates and these results lead to competitive 
confidence intervals for quantiles, simple ratios of two quantiles \citep{prst-2016b} and other measures of inequality \citep{prst-2017,prst-2018}. Further, useful applications are robust analogues to the coefficient of variation \citep{arachchige2022robust} and the fitting of the generalized lambda distributions \citep{shenal-2020}.  These aforementioned works, and others that have appeared recently, show that confidence intervals obtained using the estimates and standard errors exhibit excellent coverage (close-to-nominal) for a wide range of underlying distributions, including those that are heavily skewed.  Similarly, valid hypothesis testing is also achievable.

Despite the existence of confidence intervals for quantiles, linear combinations of quantiles, and ratios of linear combinations, functionality to obtain them are not readily available for researchers.  This is perhaps one reason why the mean and other moment-based estimators are so widely reported in the literature, even for data types for which there is evidence that this is not advised.  For example, \cite{arachchige2021interval} provided examples showing that despite very visible differences in dispersion between two independent samples in the presence of skew, this was not detected using ratios of sample variances.  On the contrary, ratios of interquantile ranges did provide evidence of differences. 

Comparisons of quantiles, and functions of quantiles, can also provide rich insights that are not available using moment-based estimators.  For example, \cite{price2021impact} consider the effect of preconceptions of weight loss on pregnancy outcomes.  Using methods from \cite{prst-2016a}, they showed that there was a significant difference in the third quartiles of birth weight centiles when comparing two different weight loss interventions.  This analysis was able to clearly show that one intervention led to birth weights that were skewed towards the upper end of the distributions of all weights.

The aim of the \texttt{rquest} package is to avail researchers with simple-to-use functionality to obtain reliable point and interval estimators, and to carry out statistical inference on measures based on quantiles.  The functionality is flexible, allowing the user to define their own ratios of linear combinations of quantiles in addition to the many well known measures that are included.  The package can be downloaded from CRAN at \url{https://cran.r-project.org/package=rquest}.  In Section 2 we detail some of the theoretical background required, before describing the package functionality in Section 3.  Examples are considered in Section 4 and we conclude in Section 5.

\section{Theory and estimation}

For a random variable $X$ with distribution function $F$ with $F(x)=P(X\leq x)$, let $Q$ denote the associated \textit{quantile function} where $x_p\equiv Q(p)=F^{-1}(p)$ is such that $P(X\leq x_p)=p$ with $p\in[0,1]$.  We also consider the \textit{quantile density function} \citep[e.g.,][]{parzen1979nonparametric} to be \begin{equation}
 q(p)=Q'(p)=\frac{1}{f(x_p)}\label{qdf}   
\end{equation} 
where $f$ is the probability density function; i.e., $f(x)=F'(x)$.  The quantile density function appears in asymptotic variances and covariances of quantile estimators, as follows.

Consider a random sample of size $n$ denoted by $X_1,\ldots,X_n$ and let $F_n$ denote the associated empirical distribution function.  An estimator of $x_p$ from this sample is $\widehat{x}_p=F_n^{-1}(p)$.  The asymptotic variance of $\widehat{x}_p$, and the asymptotic covariance between two quantile estimators $\widehat{x}_{p}$ and $\widehat{x}_{r}$ from this same sample is \cite[e.g., Chapter 7 of][]{dasgupta2008asymptotic},
\begin{equation}
    n\ \text{var}(\widehat{x}_{p})\ \dot{=}\ p(1-p)q^2(p),\;\; n\ \text{cov}(\widehat{x}_{p},\widehat{x}_{r}) \ \dot{=}\  p(1-r)q(p)q(r)\label{eq:var_and_cov}
\end{equation}
for $p < r$.  Hence, given estimates of the quantile density function, standard errors and estimates of the covariance between two dependent quantile estimators can be found leading to confidence intervals and hypothesis testing.

\subsection{Some measures of location, spread, skewness and inequality based on quantiles}

Given a sample of $n$ observations, $x_1,\ldots,x_n$, let $\widehat{x}_p=\widehat{Q(p)}$ denote the sample $p$th quantile (i.e. the estimate of $Q(p)$). For many measures based on quantiles, it is straightforward to obtain estimates of these measure by replacing the population quantiles with their estimates.  We detail many of these measures below, and discuss other aspects of estimation when it is necessary to do so.

\subsubsection{Location}

As quantile measures of location, the median is most commonly reported, follow by the first, and third quartiles, $Q(0.25)$ and $Q(0.75)$.  The median is often found when describing skewed data (e.g., houseprices, incomes etc.), since it is a better indicator of `typical', or of `centrality', when compared to the mean.  There continues to be advocacy for wider spread usage of the median.  For example, with respect to pharmacology, \cite{marino2014use} highlight that the mean is not always a good choice, and that ``the median provides an excellent summary statistic".

\subsubsection{Dispersion}

When dealing with skewed distributions, a commonly reported measure of spread is the interquartile range, $\text{IQR}=Q(0.75)-Q(0.25)$, the difference between the third and first `quartiles'.  When comparing dispersion between skewed data sets, or data sets with outliers, the comparison of IQRs can be vastly superior.  For example, data shapes can be very different when it comes to spread (as depicted by boxplots and histograms), and yet this may not be evident when comparing sample variances.  \cite{arachchige2021interval} provide real examples of where outliers and/or skew negatively impact sample variances to the point where data dispersion seems to be similar, and yet IQRs indicate very large differences.

The coefficient of variation (CV), defined as the standard deviation divided by the mean, is a popular measure of relative dispersion.  \cite{arachchige2022robust} show that for skewed distributions and/or samples including outliers, the CV may not be an appropriate measure of relative dispersion.  Instead, they provide examples and confidence intervals for the robust CV measure given as
\begin{equation}
    \text{rCV} = 0.75\times \frac{IQR}{M}\label{rCV}
\end{equation}
where the multiplier $0.75$ is chosen so that CV and rCV are approximately equivalent for the Gaussian distribution.

\subsubsection{Skewness and kurtosis}

Bowley's measure of skew is the difference between the distance of the third quartile from the median $M=Q(0.5)$ and the distance between the median and the first quartile, relative to the IQR.  That is,
\begin{equation}
B = \frac{[Q(0.75) - M] - [M - Q(0.25)]}{\text{IQR}}=\frac{Q(0.75) + Q(0.25) - 2M}{\text{IQR}}.\label{B}
\end{equation}

It is possible to generalize Bowley's skew \citep{groeneveld1984measuring}, so that the measure can focus on either points further out in the tails, or closer to the median.  The generalizaton is, for $p\in (0,1/2)$,
\begin{equation}
B(p) = =\frac{Q(1-p) + Q(p) - 2M}{\text{IQR}}.\label{Bp}
\end{equation}
where $B(0.25)$ is equal to Bowley's skew in \eqref{B}.  The measure based on the outer deciles, e.g., $B(0.1)$ is commonly referred to as Kelly's skewness measure.

Additionally, there have been further variations of the measure.  For example, \cite{groeneveld2009improved} provide a measure of skew that can be used when the direction of skew is known.  For right skew, this measure is
\begin{equation}
    G(p)=\frac{Q(1-p) + Q(p) - 2M}{M - Q(p)}\label{Gp}
\end{equation}
and for left skew the denominator is $Q(1-p)-M$.

There also exist quantile-based measures of kurtosis to measure weight in the tails of a distribution and peakedness.  For further details on these concepts based on quantiles measures, including inference, see \cite{staudte2017inference}.  The kurtosis measure by \cite{moors1988quantile} is given as
\begin{equation}
    T=\frac{Q(7/8)-Q(5/8) + Q(3/8) - Q(1/8)}{Q(6/8)-Q(2/8)}.\label{moors}
\end{equation}
Other options for measuring kurtosis including focusing on the left and right tail weights specifically.  For example, \cite{brys2006robust} introduce the right and left measures of tail weight defined respectively to be, for $p\in (0,1/2)$ and $q\in (0.5, 1)$,
\begin{equation}
    \text{LQW}(p)=\frac{Q[(1-p)/2]+Q(p/2)-2Q(0.25)}{Q[(1-p)/2]-Q(p/2)},\;\; \text{RQW}(q)=\frac{Q[(1+q)/2]+Q(1-q/2)-2Q(0.75)}{Q[(1+q)/2]-Q(1-q/2)}.\label{tail_weight}
\end{equation}

\subsubsection{Inequality}\label{sec:inequality_measures}

Symmetric ratios of quantiles, e.g. $Q(1-p)/Q(p)$, have been a popular measure of inequality.  \cite{prendergast2017large} introduce confidence intervals for these ratios, and also non-symmetric ratios, highlighting the close-to-nominal coverage that can be achieved.  More broadly, inequality measures based on quantile ratios are becoming increasingly popular and the advantages of these compared to those based on moment estimators are well described by \cite{brazauskas2024measuring}.

\cite{prst-2018} use the idea of ratios of quantiles for assessing inequality to introduce the inequality measure, called the \textit{Quantile Ratio Index (QRI)},
\begin{equation*}
    \text{QRI} = 1 - \int^1_0\frac{Q(p/2)}{Q(1-p/2)}dp
\end{equation*}
which is equal to zero only if all populations values are equal, and with greater QRI indicating greater inequality.  For a suitably large positive integer $J$ (e.g. $J=100$) and $p_i=(i-0.5)/J$ for $i=1,\ldots,J$, QRI can be estimated by
\begin{equation}
    \widehat{\text{QRI}}_J =\frac{1}{J}\sum^n_{i=1}\left[1-\frac{\widehat{x}_{p_i/2}}{\widehat{x}_{1-p_i/2}}\right].\label{I_hat}
\end{equation}

Prior to introducing the QRI, \cite{quantle_lorenz} considered quantile versions of the Lorenz curve.  The Lorenz curve depicts inequality and is typically used to describe inequality amongst incomes.  The Gini index \citep[$G$, ][]{gini-1914}, is the proportion of area between the line of perfect equality and the Lorenz curve, relative to the area below line of equality.  Consequently, and similar to interpretation of QRI, a $G$ of zero indicates perfect equality and increasing values up to one represents increasing inequality.  For more on the Lorenz curve and $G$ see, for example, \cite{kleiber2008lorenz}.

The proportional area between inequality and one of the quantile-based Lorzenz curves considered by \cite{quantle_lorenz} that is based on symmetric ratio of quantiles, leads to a quantile variant of $G$.  This inequality measure is
\begin{equation}
    G_2=2\int^1_0p\left[1-\frac{Q(p/2)}{Q(1-p/2)}\right]dp=1-2\int^1_0p\frac{Q(p/2)}{Q(1-p/2)}dp.
\end{equation}
sharing similarities with QRI.  Similarly to the estimation of QRI in \eqref{I_hat}, $G_2$ can be estimated approximately based on the ratio of symmetric quantiles using $p_1,\ldots,p_J$ defined above.

\subsection{Estimation of the quantile density function and approximate variances of quantile estimators}\label{sec:estimation_of_pdq}

Recall the definition of the quantile definition function $q(p)=1/f(x_p)$ in Equation \ref{qdf}.  While it is possible to estimate $q(u)$ by taking the inverse of a density estimator evaluated at $\widehat{x}_u$, it is usually preferable to estimate $q(u)$ directly \citep[e.g.][]{jones-1992}.  For an ordered random sample $X_{(1)}<X_{(2)}< \ldots < X_{(n)}$ and kernel function, $K$, a direct estimator of $q(p)$ is
\begin{equation}
    \widehat{q}(p)=\sum^n_{i=1}X_{(i)}\left[K_b\left(p - \frac{i-1}{n}\right)-K_b\left(p - \frac{i}{n}\right)\right]\label{q_hat}
\end{equation}
where $K_b(y)=K(y/b)/b$ for a suitably chosen bandwidth $b$.  \cite{jones-1992} provides the mean square error (MSE) of $\widehat{q(p)}$ and the minimum of this MSE depends on the bandwidth, $b$.  The optimal $b$ can be determined via the \textit{quantile optimality ratio} (QOR), defined as
\begin{equation}
\text{QOR}(p)=\frac{q(p)}{q''(p)}.\label{QOR}
\end{equation}

\cite{prst-2016a} provide methods for estimating the QOR, leading to an estimate of the optimal $b$ and estimation of $q(p)$ shown in \eqref{q_hat}.  One option considered is to estimate the density function, $f$, using the very flexible four parameter Generalised Lambda Distribution \cite[GLD, for which they use the so called FKML parameterization][]{freimer1988study}.  While this leads to good estimates of $q(p)$ and consequently good estimates of the asymptotic variances and covariances for quantile estimators (see Equation \ref{eq:var_and_cov}, it is also shown that across many different underlying distributions, very good results can also be achieved by using the estimated QOR for the lognormal distribution.  The approach leads to lesser computation times due to estimating just two parameters.

Let $\mathbf{p}=[p_1,\ldots,p_d]^\top$ be a vector of length $d$ with elements in $(0,1)$ indicating the quantiles to be estimated.  We consider the $d\times d$ symmetric matrix, $\widehat{\boldsymbol{\Sigma}}_\mathbf{p}$, of estimated variances and covariances for these quantile estimators.  Hence, the $j$th diagonal element is the estimate to $\text{var}(\widehat{x}_{p_j})$ and the $ij$th $(i\neq j)$ is the estimate to $\text{cov}(\widehat{x}_{p_i},\widehat{x}_{p_j})$.  Let $\mathbf{b}_1\in \mathcal{R}^d$ and $\mathbf{b}_2\in \mathcal{R}^d$ be vectors of quantile coefficient that will define the linear combinations of the quantiles to be taken and let $\widehat{\mathbf{x}}_\mathbf{p}=[\widehat{x}_{p_1},\ldots,\widehat{x}_{p_d}]^\top$ be the vector of quantile estimates.  Then we have the following estimates for variances of the linear combinations of quantiles and covariance between these combinations as
\begin{equation}
  v_1 = \widehat{\text{var}}(\mathbf{b}_1^\top \widehat{\mathbf{x}}_\mathbf{p})=\mathbf{b}_1^\top\widehat{\boldsymbol{\Sigma}}_\mathbf{p}\mathbf{b}_1 ,\; v_2 = \widehat{\text{var}}(\mathbf{b}_2^\top \widehat{\mathbf{x}}_\mathbf{p})=\mathbf{b}_2^\top\widehat{\boldsymbol{\Sigma}}_\mathbf{p}\mathbf{b}_2 \;\text{and}\;v_{12} = \widehat{\text{cov}}(\mathbf{b}_1^\top \widehat{\mathbf{x}}_\mathbf{p},\mathbf{b}_2^\top \widehat{\mathbf{x}}_\mathbf{p})=\mathbf{b}_1^\top\widehat{\boldsymbol{\Sigma}}_\mathbf{p}\mathbf{b}_2.\label{eq:var_lin_comb} 
\end{equation}
Finally, we can use the Delta method \cite[e.g. Chapter 11 of ][]{dasgupta2008asymptotic} to obtain estimated variances for the ratio of linear combinations of quantiles and the log of the ratio of linear combinations.  Let $R=\mathbf{b}_1^\top \widehat{\mathbf{x}}_\mathbf{p}/\mathbf{b}_2^\top \widehat{\mathbf{x}}_\mathbf{p}$.  Then
\begin{equation}
\widehat{\text{var}}(R)=R^2\left[\frac{v_1}{(\mathbf{b}_1^\top \widehat{\mathbf{x}}_\mathbf{p})^2}+\frac{v_2}{(\mathbf{b}_2^\top \widehat{\mathbf{x}}_\mathbf{p})^2}-2\frac{v_{12}}{\mathbf{b}_1^\top \widehat{\mathbf{x}}_\mathbf{p}\mathbf{b}_2^\top \widehat{\mathbf{x}}_\mathbf{p}}\right]\label{eq:var_R}
\end{equation}
where $v_1,v_2,v_{12}$ are given in \eqref{eq:var_and_cov} and
\begin{equation}
    \widehat{\text{var}}[\log(R)]=\frac{1}{R^2}\widehat{\text{var}}(R).\label{eq:var_log_R}
\end{equation}
We provide this approximate variance since the log of a ratio generally has superior statistical properties.

\subsection{Confidence intervals and hypothesis testing}

In recent years, there has been much research showing that asymptotic Wald-type confidence intervals for quantiles, ratios of quantiles and ratios of linear combinations of quantiles using the variances in Section \ref{sec:estimation_of_pdq} can achieve good coverage for not-too-small samples sizes (e.g. $n\geq 30$).  Therefore the confidence intervals for obtaining them in \texttt{rquest} are of this form.  There are other options for estimating confidence intervals and some of these are also available in R.  For example, the package \texttt{Qtools} \citep{geraci2016qtools} provides functionality for obtaining confidence intervals for single quantiles and this approach may be preferred for smaller sample sizes.  Other packages that can be used to obtain intervals for single quantiles include \texttt{QuantileNPCI} \citep{QuantileNPCI} and \texttt{DescTools} \citep{DescTools}.

As an example for the Wald intervals, the $(1-\alpha/2)\times 100$\% confidence interval for the robust CV can be obtained in two ways.  Let $\mathbf{p}=[0.25,0.5,0.75]^\top$, $\widehat{\mathbf{x}}_{\mathbf{p}}=[\widehat{x}_{0.25}, \widehat{x}_{0.5}, \widehat{x}_{0.75}]^\top$, $\mathbf{b}_1=0.75\times [-1,0,1]^\top$, $\mathbf{b}_2=[0,1,0]^\top$ and the estimated robust CV be $$\widehat{\text{rCV}}=\frac{\mathbf{b}_1^\top \widehat{\mathbf{x}}_\mathbf{p}}{\mathbf{b}_2^\top \widehat{\mathbf{x}}_\mathbf{p}}.$$ Then, firstly, we have the interval
\begin{equation}
    \widehat{\text{rCV}}\pm z_{1-\alpha/2}\times \text{SE}
\end{equation}
where SE is the standard error which is the square root of the estimated variance calculated using \eqref{eq:var_R} and $z_{1-\alpha/2}$ is the $1-\alpha/2$ quantile from the standard normal distribution.

However, given that the log of ratios commonly results in improved statistical properties, a typically preferred approach would be to calculate the interval for the log ratio, and then to exponentiate back to the ratio scale.  That is
\begin{equation}
    \exp\left\{\log\left(\widehat{\text{rCV}}\right)\pm z_{1-\alpha/2}\times \text{SE}_{\text{log}}\right\}
\end{equation}
where $\text{SE}_{\text{log}}$ is the square root of the estimated variance in \eqref{eq:var_log_R}.

We can also compute a p-value for a hypothesis test against $H_0:\text{rCV}=r$ for a chosen $r$ using the asymptotic normality of the test statistic $T=(\widehat{\text{rCV}}-r)/\text{SE}$.  For a two-sided alternative, e.g. $H_1:\text{rCV}\neq r$, this p-value can be computed as $2\times \Phi(-|T|)$ where $\Phi$ is the distribution function for the $N(0,1)$ distribution. 

For the two sample case (assuming independent samples), hypothesis testing and confidence interval estimation for differences and ratios follow naturally from what is described above.

\section{Package functionality}

In this section we discuss functionality available in the package with some instructions for the various ways in which users can obtain results for one- and two-sample analyses of quantile-based measures.

\subsection{Variances and covariances of quantile estimators}

Function \texttt{qcov()} returns a covariance matrix consisting of variances (on the diagonal) for quantile estimates and covariances (off-diagonal) between different quantile estimates.  It is of the form:
\begin{verbatim}
qcov(x, u, method = "qor", FUN = qor.ln, quantile.type = 8, bw.correct = TRUE)
\end{verbatim}
where \textit{x} is a numeric vector of data and \textit{u} is a numeric vector of probabilities specifying the quantiles to be estimated.   The default method is to estimate the probability density function directly using the lognormal QOR for choosing a suitable bandwidth (see Section \ref{sec:estimation_of_pdq} above for more details).  Alternatively, the variances can be estimated by inverting a density estimator evaluated at the quantiles and this can be done using \texttt{method = "density"}.  Our preference is to use the quantile estimators recommended by \cite{hyndman1996sample} (type 8; see help file for \texttt{quantile()} for more details). Other estimators can be used by changing the argument \textit{quantile.type}.

As a simple example, consider the sample {\it quartiles} (first quartile $Q(0.25)$, median $Q(0.5)$, and third quartile $Q(0.75)$).  Then the code below returns the variances and covariances of the estimators for a random generated sample of 100 normal observations:

\begin{verbatim}
set.seed(1234)
x <- rnorm(100)
qcov(x, c(0.25, 0.5, 0.75))   

$cov
            0.25         0.5        0.75
0.25 0.014761324 0.008562415 0.007882055
0.5  0.008562415 0.014900078 0.013716134
0.75 0.007882055 0.013716134 0.037878793
\end{verbatim}

The row and column names indicate the probability specifying the value for $u$.  For example, the variance for the median estimator is approximately, 0.0149, and the covariance between the median and the first quartile estimator is 0.0086.

\subsection{Quantiles and linear combinations of quantiles}

\begin{verbatim}
q.test()
\end{verbatim}

\noindent The \texttt{q.test()} function can be used to carry out hypothesis tests and obtain associated confidence intervals for linear combinations of quantiles, and ratios of  such linear combinations.   With respect to its most basic usage, hypothesis tests and confidence intervals for a single quantile can be obtained.  Returning a \texttt{htest} class object, output will appear similar to that of other R functionality such as the \texttt{t.test()}.

\begin{verbatim}
q.test(x, y = NULL, measure = "median", u = NULL, coef = NULL, u2 = NULL, 
  coef2 = NULL, quantile.type = 8, var.method = "qor", 
  alternative = c("two.sided","less","greater"),  conf.level = 0.95, 
  true.q = 0, log.transf = FALSE, back.transf = FALSE, min.q = -Inf, p = NULL)  
\end{verbatim}

\noindent For $M=Q(0.5)$, the median, the default usage of \texttt{q.test(x)} where \texttt{x} is a numeric vector, will return results for the test $H_0:M = 0$ versus $H_1:M \neq 0$, as specified by the default choice of \texttt{measure = median}, \texttt{argument} defaults to the first choice of \texttt{two.sided} for a two-sided alternative and the null hypothesis for which to test against for the median is equal to 0 (\texttt{true.q = 0}).  A 95\% confidence interval for $M$ will also be computed (\texttt{conf.level = 0.95}). 

There are several ways to implement \texttt{q.test} for either one or two samples.  For many well known measures, the simplest way to do this is to use the argument \texttt{measure} to specify which measure.  However, linear combinations and ratios of such linear combinations can also be defined by the user.

\subsubsection{Using argument \textit{measure}}

The default use of \texttt{q.test} is for a test of a single median, or a comparison of two medians when there are two samples.  For example, given two numeric vectors \texttt{x} and \texttt{y}, the following will carry out a test comparing two medians (null hypothesis of zero difference) and compute a confidence interval for the difference in two medians.  

\begin{verbatim}
q.test(x, y)    
\end{verbatim}

\begin{table}[ht]
\centering
\begin{tabular}{llll}\toprule
  Measure & \textit{measure}   & \textit{u} & \textit{coef}   \\ \midrule
\rowcolor[gray]{0.9} 
  median & \texttt{"median"}   & \texttt{u=0.5} & \texttt{coef=1} \\ 
  interquartile range & \texttt{"iqr"} & \texttt{u=c(0.25, 0.75)} & \texttt{coef =c(-1,1)} \\
  \rowcolor[gray]{0.9} 
 Robust CV & "rCViqr" & \texttt{u = c(0.25,0.75)} & \texttt{coef=0.75*c(-1,1)}\\
 \rowcolor[gray]{0.9} 
  & &\texttt{u2 = 0.5}  & \texttt{coef2=1}\\
  Bowley's skew & "bowley" & \texttt{u = c(p,0.5,1-p)} & \texttt{coef=c(1,-2,1)}\\
  & &\texttt{u2 = c(p,1-p)}  & \texttt{coef2=c(-1,1)}\\
  \rowcolor[gray]{0.9} 
    Kelly's skew & "kelly" & \texttt{u=c(0.1,0.5,0.9)} & \texttt{coef=c(1,-2,1)}\\
    \rowcolor[gray]{0.9} 
  & & \texttt{u2=(0.1,0.9)} & \texttt{coef2=c(-1,1)}\\
  Groeneveld \& Meeden's& "groenR" & \texttt{u=c(p,0.5,1-p)} & \texttt{coef=(1,-2,1)}\\
 \;\;right skew  & & \texttt{u2=c(p,0.5)} & \texttt{coef2=c(-1,1)}\\
 \rowcolor[gray]{0.9} 
    Groeneveld \& Meeden's & "groenL" & \texttt{u=c(p,0.5,1-p)} & \texttt{coef=c(1,-2,1)}\\
    \rowcolor[gray]{0.9} 
  \;\;left skew  & & \texttt{u2=c(0.5,1-p)} & \texttt{coef2=c(-1,1)}\\
        Moors kurtosis & "moors" & \texttt{u=c(1/8,3/8,5/8,7/8)} & \texttt{coef=c(-1,1,-1,1)}\\
   & & \texttt{u2=c(2/8,6/8)}& \texttt{coef2=c(-1,1)}\\
   \rowcolor[gray]{0.9} 
 Left quantile & "lqw" & \texttt{u=c(p/2,0.25,(1-p)/2)} & \texttt{coef=c(1, -2, 1)}\\
 \rowcolor[gray]{0.9} 
 \;\;tail weight  & & \texttt{u2=c(p/2,(1-p)/2)}& \texttt{coef2=c(-1,1)}\\
 Right quantile & "rqw" & \texttt{u=c(1-p/2,0.75,(1+p)/2)} & \texttt{coef=c(1, -2, 1)}\\
 \;\;tail weight  & & \texttt{u2=c(1-p/2,(1+p)/2)}& \texttt{coef2=c(-1,1)}\\
 \rowcolor[gray]{0.9} 
  quantile ratio (xx/yy) & \texttt{"qrxxyy"} & \texttt{u=xx} & \texttt{coef=1} \\
  \rowcolor[gray]{0.9} 
  & & \texttt{u2=yy} & \texttt{coef2=1}\\
    quantile ratio (90/10) & \texttt{"qr9010"} & \texttt{u=0.9} & \texttt{coef=1} \\
  & & \texttt{u2=0.1} & \texttt{coef2=1}\\
  \bottomrule 
\end{tabular}
\caption{Pre-defined measures that are accessible in the \texttt{q.test()} function using a character string argument for \textit{measure}.  The choices of \textit{u} and \textit{coef} (and \texttt{u2} and \texttt{coef2}) in vector or matrix form are shown that would achieve the same result.}\label{tab:usage}
\end{table}

The available options for argument \textit{measure} are shown in Table \ref{tab:usage}.  In this table, we also include the choices of the probability vector \texttt{u} (and \texttt{u2} for ratio measures) and \texttt{coef} (and \texttt{coef2}) that would achieve the same result.  However, is using \texttt{measure} (e.g. \texttt{measure == "median"}) then these are not required to be inputted by the user.  We provide some additional comments below.

\begin{itemize}
    \item \texttt{"median"}: The default choice.
    \item \texttt{"iqr"}: The interquartile range denoted $Q(0.75)-Q(0.25)$.
    \item \texttt{"rCViqr"}: Is for the robust CV measure denoted $0.75\times[Q(0.75)-Q(0.25)]/M$ \citep[see][]{arachchige2022robust}.
    \item \texttt{"bowley", "groenR" and "groenL"}: These choices are for Bowley's skew coefficient, given in \eqref{B} and \eqref{Bp} for the generalised measure, and the right and left skew measures by Groeneveld and Meeden shown in \eqref{Gp}.  For each of these measures, the user may specify the choice of $p$ using argument \texttt{p}.  If this is not used, then the default is \texttt{p=0.25}.
    \item \texttt{"moors"}: This is the choice for Moors kurtosis measure (see Eq. \ref{moors}).
    \item \texttt{"lqw" and "rqw"}: These are for the robust left and right tail weights proposed by \cite{brys2006robust} and shown in \eqref{tail_weight}.
    \item "qrxxyy": A character string consisting of the first two characters \texttt{"qr"} and followed by four numbers will request a ratio of quantiles.  The first two number digits (in place of \texttt{xx}) will indicate the quantile for the numerator, and the second two numerical digits \texttt{yy} for the denominator.  For example, \texttt{"qr9010"} will estimate and test the ratio $Q(0.9)/Q(0.1)$.
\end{itemize}

\subsubsection{Log transformation and back-transformations}

When estimating ratios, it may be preferable to use the log transformation to improve statistical properties of the estimator.  This can be achieved using \texttt{log.transf == TRUE} in the call to \texttt{q.test}.  If one wishes to return estimates (point and interval) on the ratio scale, then \texttt{back.transf == TRUE} can be used.

In the two-sample setting, the log transformation will apply log to the ratio of the two estimates (equivalent to the difference in the log of the estimates).  Again, if \texttt{back.transf == TRUE} is used, then results will be presented on the ratio scale.  One difference with the two-sample setting is that if argument \textit{true.q} is equal to zero (the default) and \texttt{back.transf == TRUE}, then \textit{true.q} will be set equal to one for testing whether the ratio is equal to one (true values equal).  This will not happen if \textit{true.q} has been set to a different value thereby overriding the test of equal values.

\subsubsection{User-defined measures using arguments \textit{u} and \textit{coef}}

If argument $\texttt{u}$ is not \texttt{NULL}, then this overrides the choice for argument \texttt{measure}. 
 Tests and intervals for other quantiles, $Q(u)$, can be obtained by choosing the appropriate value for argument \texttt{u}.  E.g., \texttt{u = 0.25} will obtain the results pertaining to the first quartile.

Linear combinations of quantiles are also possible. 
 Let $\mathbf{b}=[b_1,\ldots,b_d]^\top\in \mathcal{R}^d$ denote a vector of real valued coefficients.  Also suppose that $\mathbf{x}=[Q(u_1),\ldots,Q(u_d)]^\top$ be a vector of quantiles.  Then tests for $\mathbf{b}^\top\mathbf{x}=b_1Q(u_1)+\ldots b_dQ(u_d)$ by choosing the required values for arguments \texttt{u} and \texttt{coef}. For example, if \texttt{u = c(0.25, 0.75)} and \texttt{coef = c(-1, 1)}, then a test and interval for the interquartile range (IQR), $Q(0.75) - Q(0.25)$ will result.  Additionally, if \texttt{coef} is \texttt{NULL} then \texttt{coef} will be set equal to a vector with elements equal to one, and of equal length to the length of \texttt{u}.

Finally, ratios of linear combinations or, in its simplest form, ratios of two quantiles, can also be obtained.  Here, \texttt{u} is still a $p$-dimensional vector indicating the quantiles to be used, and \texttt{coef} is a $2\times p$ matrix whose first row specifies the coefficients for the numerator and the second row for the denominator for the quantiles obtained as indicated by argument \texttt{u}.   Alternatively, if \texttt{coef2} is not \texttt{NULL} and a vector, then this contains the coefficients for the denominator in the ratio. Then there are two usage scenarios:
\begin{itemize}
    \item if \texttt{u2} is \texttt{NULL}, then \texttt{coef2} will use the quantiles defined in \texttt{u}. In this case \texttt{coef2} needs to be the same length as \texttt{u} where a zero coefficient will indicate that the associated quantile is not used in the denominator linear combination.
    \item If \texttt{u2} is not \texttt{NULL}, then it must be the same length as \texttt{coef2} and provide the probabilities that indicate which quantiles are to be used in the linear combination for the denominator.
\end{itemize}

\subsection{Quantile inequality measures}

Quantile-based inequality measures are available either using \texttt{q.test} when ratios of quantiles or ratios of linear combinations are required, or using \texttt{qineq} when more complicated measures are needed.

\begin{verbatim}
qineq()
\end{verbatim}

\noindent The \texttt{qineq()} function can be used to carry out hypothesis tests and obtain associated confidence intervals for the QRI and $G_2$ measures detailed in Section \ref{sec:inequality_measures}.  Similarly to \texttt{q.test} function, a \texttt{htest} object will be returned.

\begin{verbatim}
qineq(x, y = NULL, J = 100, measure = "QRI", alternative = c("two.sided", 
  "less", "greater"), quantile.type = 8, var.method = "qor", conf.level = 
  0.95, true.ineq = 0.5)
\end{verbatim}

Currently, the argument \texttt{measure} will only accept \texttt{"QRI"} (the default), or \texttt{"G2"}.  However, more measures may be included later.  Many arguments are of identical name and usage as for \texttt{q.test}, and the reader is referred to the previous section for more details on those.  Arguments that do differ are:
\begin{itemize}
    \item \texttt{J}: this is the value for $J$ shown in, e.g. for estimation of QRI in \eqref{I_hat} and where the choice of 100 is default.
    \item \texttt{true.ineq}: the value of the inequality measure under the null hypothesis.  Here we have chosen a default of 0.5. 
\end{itemize}

Unlike the \texttt{qtest} function, presently it is not possible for users to define their own measures.

\section{Examples}

In this section we provide some example usage of the \texttt{q.test} and \texttt{qineq} functions. 

\subsection{One sample example: bladder cancer remission}

We will first consider a univariate example of bladder cancer remission times, in months, for 132 patients.  The data is highly skewed, and was recently fitted using the Burr III Marshal Olkin family of distributions by \cite{bhatti2019burr}.  The data is available in the \texttt{DataSetsUni} package \citep{DataSetsUni} which is our source for the analysis below.

\begin{verbatim}
library(DataSetsUni)
t.test(data_bldercancer)

	One Sample t-test

data:  data_bldercancer
t = 10.083, df = 127, p-value < 2.2e-16
alternative hypothesis: true mean is not equal to 0
95 percent confidence interval:
  7.52767 11.20358
sample estimates:
mean of x 
 9.365625 
\end{verbatim}

The mean remission time is approximately 9.4 days.  However, close to 70\% of the remission times are less than this value and so the mean should not be considered a typical remission time.  We also note that the width of the 95\% confidence interval, which is $(7.53,\ 11.20)$ for the mean is around 3.7 days.

We can easily use \texttt{q.test} to obtain a 95\% confidence interval for the median.

\begin{verbatim}
q.test(data_bldercancer)

	One sample test of the median

data:  data_bldercancer
Z = 9.408, p-value < 2.2e-16
alternative hypothesis: true median  is not equal to 0
95 percent confidence interval:
 5.062726 7.727274
sample estimates:
median  
  6.395 
\end{verbatim}

The estimated median is 6.4 and the 95\% confidence interval for the median is approximately $(5.06,\ 7.73)$.  The estimate and interval are better indicators of a typical remission time and the interval is considerably shorter than that for the mean (by more than one day).  We may also be interested in, for example, estimating the remission time for which 75\% of patients results, and do so we can set argument \texttt{u} to 0.75.

\begin{verbatim}
> q.test(data_bldercancer, u = 0.75)$conf.int
[1]  9.168747 14.632920
attr(,"conf.level")
[1] 0.95    
\end{verbatim}

\subsection{Two sample examples}\label{sec:two_sample_examples}

In this section we consider Melbourne house prices data available from \url{https://www.kaggle.com/datasets/saadmehar/melbourne-housing-fullcsv}.  The full data set consists of nearly 34,000 sale prices (AUD) from publicly available dataacross 351 suburbs.  We will consider comparing house prices for two neighboring suburbs, Bundoora and Kingsbury.  We store the values of each in \texttt{x} and \texttt{y} below and scale the prices so that they are units of one million.  In total there are 411 sales for Bundoora and 39 for Kingsbury.

\begin{verbatim}
data <- read.csv(file="HPriceKB.csv", header=TRUE, sep=",")
x <- data$HousePrice[data$Suburb == "Bundoora"]
y <- data$HousePrice[data$Suburb == "Kingsbury"]
\end{verbatim}

\begin{figure}[h!t]
    \centering
    \includegraphics[scale = 0.35]{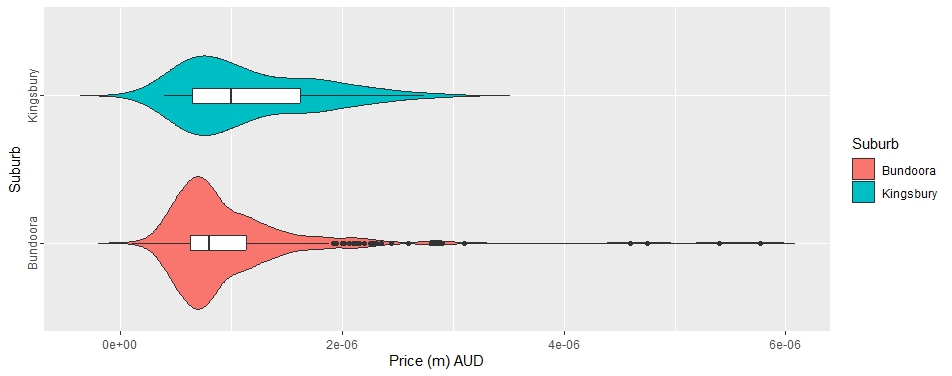}
    \caption{Violin plots of house prices for two neighboring suburbs in Melbourne, Australia.}
    \label{fig:violin_plots}
\end{figure}

Violin plots using function \texttt{geom\_violin} from the \href{https://cran.r-project.org/web/packages/ggplot2/index.html}{ggplot2} package \citep{ggplot2} are shown in Figure \ref{fig:violin_plots}.  The shape in the empirical densities differ greatly, and one would expect differences in location, spread, skewness and inequality.  The embedded boxplots also suggest these expectations.   

\subsubsection{Comparison of relative dispersion for house prices}

\cite{arachchige2022robust} introduced confidence intervals for ratios of robust coefficients of variation (robust CVs) and here we consider the robust CV defined to be $0.75\times \text{IQR}/M$ where the multiplier 0.75 is to ensure that approximate equivalence if obtained for a normal distribution.  \cite{arachchige2022robust} showed that the usual CV (SD/mean) suggests that relative spread of prices is greater for the suburb of Bundoora and we calculate these CVs below. 

\begin{verbatim}
# Bundoora CV
sd(x)/mean(x) 
[1] 0.6511846

# Kinsgury CV
sd(y)/mean(y)
[1] 0.5095749
\end{verbatim}

As we can see, for Bundoora the relative dispersion according to the CV measure is greater than the relative dispersion for Kinsbury (0.6512 $>$ 0.5096).   However, as suggested by the violin plots, this appears only due to a small number of outlying prices, with relative spread greater for Kinsgbury when those outliers are ignored.  To replicate the results we use the log transformation and back-transform to the ratio scale:

\begin{verbatim}
q.test(x, y, measure = "rCViqr", log.transf = TRUE, back.transf = TRUE)

	Two sample test of the robust coefficient of variation (0.75*IQR/median)

data:  x and y
Z = -2.0898, p-value = 0.03664
alternative hypothesis: true ratio of Robust CVs  is not equal to 1
95 percent confidence interval:
 0.4335736 0.9735627
sample estimates:
ratio of Robust CVs  
           0.6497008 
\end{verbatim}

The estimated ratio of the robust CVs comparing the two suburbs and associated 95\% confidence interval of $(0.434, 0.974)$ indicate a significantly smaller relative spread in prices in Bundoora (the numerator for the ratio of robust CVs).

In the above, we used the log transformation since the robust CV is a ratio measure.  We then back-transformed back to the ratio scale, and in doing so the comparison is for the ratio of robust CVs.  We can also choose not use the log-transformation at all, in which case we estimate the difference between the robust IQRs directly.  However, we have included a warning message to indicate that the user may wish to use the log transformation when dealing with ratio measures.  An example of this is given below.

\begin{verbatim}
q.test(x, y, measure = "rCViqr")

	Two sample test of the robust coefficient of variation (0.75*IQR/median)

data:  x and y
Z = -1.7653, p-value = 0.07752
alternative hypothesis: true difference in Robust CVs  is not equal to 0
95 percent confidence interval:
 -0.54056151  0.02824889
sample estimates:
difference in Robust CVs  
               -0.2561563 

Warning message:
In q.test(x, y, measure = "rCViqr") : You may wish to consider using a 
log transformation for ratios (e.g. log.transf == TRUE).  If you choose 
to use a log transformation you can also back-transform to the ratio 
scale using (back.transf = TRUE) if you wish.    
\end{verbatim}

\subsubsection{An example using user-defined measures}

We will now consider how a user may define their own measures for testing.  To illustrate, we will replicate the comparison of robust CVs shown in the previous section.

\begin{table}[h!t]
\centering
    \begin{tabular}{lll}\toprule
    Method & Code & Output (95\% CI) \\ \midrule
    \rowcolor[gray]{0.9} 
    1     & \texttt{q.test(x,y,measure="rCViqr",log.transf=TRUE,} & \texttt{0.4335736 0.9735627} \\
    \rowcolor[gray]{0.9} 
    & \;\;\;\;\;\;\texttt{back.transf=TRUE)\$conf.int} & \\ 
    2 & \texttt{u<-c(0.25,0.75)} & \texttt{0.4335736 0.9735627} \\
    & \texttt{coef<-0.75*c(-1,1)} & \texttt{} \\
    & \texttt{u2<-0.5} & \texttt{} \\
    & \texttt{coef2<-1} & \texttt{} \\
    & \texttt{q.test(x,y,u=u,u2=u2,coef=coef,coef2=coef2,} & \texttt{} \\
    & \;\;\;\;\;\;\texttt{log.transf=TRUE,back.transf=TRUE)\$conf.int} & \texttt{} \\
    \rowcolor[gray]{0.9} 
    3 & \texttt{u<-c(0.25,0.5,0.75)} & \texttt{0.4335736 0.9735627} \\
    \rowcolor[gray]{0.9} 
    & \texttt{num <- 0.75*c(-1,0,1)} & \texttt{} \\
    \rowcolor[gray]{0.9} 
    & \texttt{den <- c(0,1,0)} & \texttt{} \\
    \rowcolor[gray]{0.9} 
    & \texttt{coef <- rbind(num, den)} & \texttt{} \\
    \rowcolor[gray]{0.9} 
    & \texttt{q.test(x,y,u=u,coef=coef,log.transf=TRUE,} & \texttt{} \\
    \rowcolor[gray]{0.9} 
    & \;\;\;\;\;\;\texttt{back.transf=TRUE)\$conf.int} & \texttt{} \\
    \bottomrule
    \end{tabular}
    \caption{Three examples of code segments that produce.  All three produce the 95\% confidence interval for the ratio of robust CVs considered in the previous section and where a log transformation (and back-transformation) has been used.}
    \label{tab:three_examples}
\end{table}

In Table \ref{tab:three_examples} we provide three different segments of code that produce the same output, in this case a 95\% confidence interval for the ratio of robust CVs.  Method 1 uses argument \texttt{measure="rCViqr"} to acheive this, and this was the approach used in Section \ref{sec:two_sample_examples}.  The output is shown in the third column.  For Method 2, we provide the probabilities for the quantiles to be used in the numerator of the measure, i.e. \texttt{u<-c(0.25,0.75)}, since the numerator is the IQR, and the coefficients for these quantiles.  Recall that the robust CV is multiplied by 0.75 so that it is approximately equal to the usual CV for the normal distribution.  Hence, the numerator is an estimate to $0.75[Q(0.75)-Q(0.25)]$ so that our coefficient for $Q(0.25)$ is $-0.75$ and for $Q(0.75)$ it is $0.75$.  These coefficients have been chosen using \texttt{coef<-0.75*c(-1,1)}.  We use \texttt{u2} and \texttt{coef2} to select the probabilities and coefficients for the denominator.  The denominator for the robust CV is simply the median, and so we use \texttt{u2=0.5} and \texttt{coef2=1} respectively. 

For Method 3, we use \texttt{u<-c(0.25,0.5,0.75)} to specify which quantiles are to be used, irrespective of whether they are for the numerator or denominator.  We then create and use a matrix object of the form
\begin{equation*}
    \left[\begin{array}{ccc}  -0.75 & 0 & 0.75 \\ 0 & 1 & 0 \end{array}\right]
\end{equation*}
where the first row indicates the coefficients for the numerator linear combination, and the second row for the denominator.  For example, only the median is used in the denominator and so the coefficients for $Q(0.25)$ and $Q(0.75)$ are zero.  These examples show that it is simple for users to define a measure that is not included in the package by using Methods 2 and 3.

\subsubsection{A quantile-based inequality comparison of house prices}

We now provide an example of how quantile-based inequality measures can provide different insights into inequality compared to moment-based measures such as the Gini index.  We will continue to use our neighbouring suburbs data from the previous sections.

We start by comparing the Gini indices for the two suburbs where we calculate the index using the \texttt{ineq} package \citep{ineq}  and complement this with a two-sample bootstrap p-value using the \texttt{compare.two.sample} function in the \texttt{FertBoot} package \citep{FertBoot}.

\begin{verbatim}
library(ineq)
ineq(x) - ineq(y)
[1] 0.01623051

library(FertBoot)
set.seed(123)
boot.res <- compare.two.sample(x, y, fun = ineq, R = 1000)
boot.res$p.value
[1] 0.374
\end{verbatim}

Despite the different density shapes depicted in Figure \ref{fig:violin_plots}, there is only a small to moderate difference in the indices between house prices for the two suburbs.  Using the bootstrapping approach, this was not significant (p-value$=0.374$).  

We now consider the quantile-based inequality measure, QRI, detailed in \eqref{sec:inequality_measures}.
\begin{verbatim}
qineq(x, y)

	Two sample test of the QRI

data:  x and y
Z = -2.7952, p-value = 0.005187
alternative hypothesis: true difference in QRI is not equal to 0
95 percent confidence interval:
 -0.09487684 -0.01666514
sample estimates:
difference in QRI 
      -0.05577099 
\end{verbatim}

While the difference in inequality between the two samples was not significant for the Gini coefficient, a significant difference was found using the QRI (95\% CI $[-0.095,\ -0.017]$).  Further, while the difference in Gini coefficient was small and positive, the magnitude of the difference in QRIs is not only three times the size, but it is also negative.  Based on the findings by \cite{prst-2018}, we suspect that the Gini coefficient is being highly influenced by a small number of large outliers which obscures the measuring of inequality for a large majority of the data.

\section{Summary}

The \texttt{rquest} package provides convenient functionality for researchers to carry out hypothesis tests and obtain confidence intervals for measures based on quantiles.  While these are based on asymptotic normality of the measures, many recent studies have shown that even moderate sample sizes (e.g. $n>30$) will produce reliable results.  We hope that the convenient usage, including the ability for the user to define their own measures for ratios of linear combinations of quantiles, results in more than simply reporting a point estimate of quantiles as is common.  Additionally, and perhaps more importantly, we hope that this encourages researchers to report measures based on quantiles when it is more appropriate to do so (e.g. for skewed data), rather than the mean and standard deviations.

\bibliographystyle{authordate4}  
\bibliography{refs}

\end{document}